\input harvmac 
\input epsf 
\def\p{\partial} 
\def\ap{\alpha'} 
\def\half{{1\over 2}}



\Title{}{\vbox{\centerline{Noncommutative Inflation and the CMB Multipoles}
}}

\centerline{Qing Guo Huang and Miao Li} 
\centerline{\it Institute of Theoretical Physics} 
\centerline{\it Academia Sinica, P.O. Box 2735} 
\centerline{\it Beijing 100080}  
\medskip 
\centerline{\tt huangqg@itp.ac.cn} 
\centerline{\tt mli@itp.ac.cn} 
 
\bigskip 

The first year results of WMAP tentatively indicate running of the spectral index 
as well as a deficit of power in the low multipoles in the CMB spectrum. The former 
can be rather easily understood in the noncommutative inflation model, and
the latter, as we shall show in this paper, still appears to be an anomaly,
even though the noncommutative inflation model already suppresses the low
multipoles to a certain degree. By fitting the power spectrum, we determine
the string scale to be $l_s\sim 4\times 10^{-29}$cm.

\Date{August, 2003} 

\nref\wmap{C. L. Bennett et al., Astrophys. J. 464, L1 (1996), 
astro-ph/9601067; 
C. L. Bennett et al., astro-ph/0302207; 
D. N. Spergel et al., astro-ph/0302209; 
L. Verde et al., astro-ph/0301218; 
H. V. Peris et al., astro-ph/0302225;
E. Komatsu et al., astro-ph/0302223.  }
\nref\bh{R. Brandenberger, P. M. Ho, Phys. Rev. D 66 (2002) 023517, 
hep-th/0203119.}
\nref\hl{Q. G. Huang, Miao Li, JHEP 0306 (2003) 014, hep-th/0304203.}
\nref\tmb{S. Tsujikawa, R. Maartens and R. Brandenberger, astro-ph/0308169. }
\nref\wt{P. Mukherjee, Y. Wang, astro-ph/0303211; 
S. L. Bridle, A. M. Lewis, J. Weller and G. Efstathiou, astro-ph/0302306; 
G. Efstathiou, astro-ph/0306431. }
\nref\ge{G. Efstathiou, astro-ph/0303127. }
\nref\cpkl{C. R. Contaldi, M. Peloso, L. Kofman and A. Linde,
astro-ph/0303636. }
\nref\ccl{J. M. Cline, R. Crotty and Julien Lesgourgues,
astro-ph/0304558. }
\nref\fz{B. Feng, X. Zhang, astro-ph/0305020. }
\nref\dcs{S. DeDeo, R. R. Caldwell and P. J. Steinhardt, 
Phys. Rev. D 67 (2003) 103509. }
\nref\ssn{M. Fukuma, Y. Kono and A. Miwa, hep-th/0307029. }
\nref\top{A. D. Oliveira-Costa, M. Tegmark, M. Zaldarriaga and A. Hamilton, 
astro-ph/0307282. }
\nref\close{A. Lasenby and C. Doran, astro-ph/0307311. } 
\nref\morelow{Y.-P. Jing and L.-Z. Fang, astro-ph/9409072;
J. Yokoyama, Phys. Rev. {\bf D59} (1999) 107303; 
E. Gaztanaga, J. Wagg, T. Multamaki, A. Montana and D.H.Hughes,
astro-ph/0304178; M. Kesden, M. Kamionkowski and A. Cooray,  astro-ph/0306597.} 
\nref\ncst{T. Yoneya, in " Wandering in the Fields ", eds. K. Kawarabayashi, 
A. Ukawa ( World Scientific, 1987), p. 419; 
M. Li and T. Yoneya, Phys. Rev. Lett. 78 (1977) 1219, hep-th/9611072; 
T. Yoneya, Prog. Theor. Phys. 103, 1081 (2000), hep-th/0004074; 
J. Polchinski, " String Theory " volume 2.}
\nref\cgg{C-S. Chu, B. R. Greene and G. Shiu, hep-th/0011241; 
F. Lizzi, G. Mangano, G. Miele and M. Peloso, hep-th/0203099. }
\nref\rsi{B. Feng, M. Li, R. J. Zhang, X. Zhang, astro-ph/0302479; 
J. E. Lidsey and R. Tavakol, astro-ph/0304113; 
M. Kawasaki, M. Yamaguchi and J. Yokoyama, hep-ph/0304161; 
L. Pogosian, S. H. H. Tye, I. Wasserman and M. Wyman, hep-th/0304188; 
E. Keski-Vakkuri, M. S. Sloth, hep-th/0306070; 
S. Cremonini, hep-th/0305244; 
K. I. Izawa, hep-ph/0305286; 
E. D. Grezia, G. Esposito, A. Funel, G. Mangano and G. Miele, gr-qc/0305050; 
D. J. H. Chung, G. Shiu, M. Trodden, astro-ph/0305193; 
M. Yamaguchi, J. Yokoyama, hep-ph/0307373. }

The first year WMAP observations already provided us unprecedentedly
accurate data on the CMB power spectrum, $C_l$ has been measured from 
multipoles $ l = 2$ to $l \sim 600$ \wmap. 
The observed temperature power spectrum is in striking agreement 
with the predictions of the inflationary $\Lambda$CDM 
cosmology, and with observations made prior to WMAP. 

Combined with other experiments, the data tentatively suggest a blue
spectrum for low $k$, and a running spectral index. If true, it is
quite hard to understand this within the framework of the usual field theoretic
inflation
models. We showed in a previous paper \hl\ that the running of the spectral
index is rather easily understood in the noncommutative inflation model
\bh. This result was expanded and reanalyzed in \tmb.

The WMAP data also show anomalously small multipole moments compared to that 
predicted by the standard $\Lambda CDM$ model \refs{\wmap,\wt}, confirming the 
observed results of COBE. 
The analysis in \wt\ suggest that the WMAP data favor a truncation at 
$k_c \sim 3 \times 10^{-4} Mpc^{-1}$, thus favoring new physics, even 
though a model with $k_c = 0$ is not strongly excluded. 
Many theorists have proposed various mechanisms to 
explain the suppression of the lower multipoles in the CMB power 
spectrum anisotropies \refs{\ge-\close}. For some earlier work and related
work, see \morelow.

In the noncommutative inflation model with a power-law inflation, a 
macroscopic scale, resulting from a combination of two microscopic scales:
the string scale and the scale governing the inflation, appears and is
quite close to the current Hubble scale. This recommends itself as a possible
indication of new phenomena appearing at this scale. We were motivated by this fact
to investigate whether there is a relation between this scale and the suppression
of low multipoles. Unfortunately, at least within this power law inflation,
the noncommutative inflation cannot account for the suppression.

The low quadrupole seen by WMAP is suggestive of unsuspected new physics that 
plays a role at large angular scales, if it is not due to the cosmic variance. 
However, WMAP also hints that the 
standard assumption of an underlying spectrum described by a constant 
index $n_s$ may be too simplistic \wmap. In fact, the large values of $dn_s/ d\ln k$ 
of WMAP rule out many
models of inflation, this is a very welcome development indeed. 
Many authors have tried to explain the large running 
of the spectral index with different models in the last few months \refs{\hl,
\rsi}. Of course, only time will tell us whether the large running of the spectral
index is not a fake.

We shall in this paper develop further the work \refs{\bh, \hl}, we will 
determine the string scale by normalizing the power spectrum in the noncommutative 
power-law inflation model. Our result is $l_s=3.86 \times 10^{-29}$cm, slightly
different 
from that  in \tmb. We also try to fit the whole power spectrum, and find that 
the low multipoles are suppressed compared to inflation without noncommutative
spacetime, but not enough to account for the data.

Spacetime is noncommutative in string/M theory \ncst, satisfying relation
\eqn\ustc{ \bigtriangleup t \bigtriangleup x \geq l^2_s,}
where $t$ and $x$ are the physical time and space. 
It is still hard to realize manifestly this relation in string theory. 
On a general ground, if inflation probes 
physics at a scale close to string scale or a related scale, one expects 
that spacetime uncertainty must have effects in the CMB power spectrum \cgg.

The spacetime effects can be expressed by a new product, star 
product, replacing the usual algebraic product. This product does not affect the 
evolution of the homogeneous background. Let $u_k$ be the Fourier mode
of the scalar perturbation. With certain assumptions, 
the authors of \bh\ arrived at the following action 
\eqn\ap{S = V \int d \eta d^3 k \half z^2_k(\eta)
(u'_{-k} u'_{k} - k^2 u_{-k} u_{k}),}
where
\eqn\scaf{\eqalign{z_k^2(\eta)&=z(\eta)^2y_k^2(\eta), \quad
y_k^2=(\beta_k^+\beta_k^-)^{\half},\cr
{d\eta \over d\tau}&=\left({\beta_k^-\over \beta_k^+}\right)^\half,\quad
\beta_k^\pm =\half (a^{\pm 2}(\tau+\l_s^2k)+a^{\pm 2}(\tau-l_s^2k)),}}
where $a(\tau)$ is the cosmic scale factor,  
the primes denote derivatives with respect to the modified conformal time $\eta$, 
as defined in the above equation, and 
$z = a {\dot \phi \over H}$. Finally, $\tau$ is the time appearing in the following
metric
\eqn\met{ds^2 = - a^{-2} d\tau^2 + a^2 d {\vec x}^2.}
Note that the star product induces shifts $\pm l_s^2k$ in time, since $1/k$ is
the uncertainty in space.  

In this paper we investigate only the power-law inflation driven by a single 
scalar field with an exponential potential 
$V(\phi) = V_0 \exp ( - \sqrt{2 / n} \phi / M_p) $. The scale factor reads
$a(t) = (\tau / l)^{n \over n+1} = ({t \over (n+1) l})^n$, and
the Hubble constant is $H = n/t$ so
$ z = a \sqrt{2/n} M_p$, where $M_p$ is the reduced Planck mass,
$l$ is a parameter  
which we will determine by normalizing the amplitude of the power spectrum.
The power spectrum of the scalar curvature perturbation is computed in \bh\
and reads
\eqn\power{P(k) = {k^2 \over 4 \pi^2 z_k^2(\eta)} 
= {n \over 8 \pi^2}{k^2 l^2_p \over a^2(\tau) y^2(\tau,k)},} 
where $l_p=M_p^{-1}$, $\tau=\tau(k)$ is the time when the fluctuation mode $k$ 
crosses the Hubble radius or when the the fluctuation mode $k$ was 
created outside the Hubble radius.

In the noncommutative case, due to saturation of
the spacetime uncertainty relation, a sufficiently IR mode k is generated outside 
the Hubble radius. As discussed in \bh, when 
\eqn\satu{k=l_s^{-1}a_{eff}, \quad a_{eff}=(\beta^+_k/\beta^-_k)^{{1\over 4}},}
the corresponding mode saturates the spacetime uncertainty, and was
created outside the horizon. Thus, the creation time is determined by the
above relation. This relation can hardly be exact, but for now we
simply adopt it.

For large $k$, the crossing horizon time is defined through
\eqn\cross{{z_k''\over z_k}=k^2,}
generalizing the usual relation, the prime in the above equation
denotes derivative with respect to $\eta$.

Denote the wave number satisfying both \satu\ and \cross\ by $k_t$, then 
we call $k<k_t$ an IR mode and $k>k_t$ a UV mode. As we shall see, numerically
$k_t$ is quite close to $H_0^{-1}$. Actually, it is slightly larger
than $H_0^{-1}$, so an IR mode just entered our horizon not long ago.
This fact is quite suggestive, and may be related to the problem of the low 
multipole power.

The mode $k$ was generated 
outside the Hubble radius if $k < k_t$ and the creation time is 
\eqn\create{\tau (k) = k l_s^2 
\left(1 + \left({k \over k_s}\right)^{2/n}\right)^{1/2},}
where $k_s = l_s^{-1} (l_s/l)^n$, and is
related to $k_c$, see eq.(12). $k_c$ is another macroscopic scale
\hl, this length scale is much greater than the current Hubble scale.
Both $n$ and $k_c$ were determined in \hl\ using the data of the
spectral index and its running. $l$, on the other hand, can be determined
only by normalizing the power spectrum, and is first done in \tmb.

If $k \ll k_s$, we have 
\eqn\screate{\tau (k) = k l_s^2 
\left(1 + \half \left({k \over k_s}\right)^{2/n}\right).}
The power spectrum in this case becomes
\eqn\smk{P(k) = A k^{4 \over n+1} 
\left(1-{5 \over 4}{n \over n+1} \left({k \over k_s}\right)^{2 \over n}\right),}
where
\eqn\smka{A = 2^{-{n-1 \over n+1}}
{n \over 8 \pi^2} \left({l \over \l_s}\right)^{4n \over n+1}
\left({l_p \over l_s}\right)^2 l_s^{4 \over n+1}.} 
Equation \smk\ tells us that 
the spectral index is $n_s = 1 + 4/(n+1)$, so in the very IR end we have a blue 
spectrum. 

For a UV mode $k>k_t$ and $k \gg k_c \sim k_s$, 
where 
\eqn\kc{k_c = \left( {n(2n-1) \over (n+1)^2} \right)^{n+1 \over 4} k_s,} 
as defined in \hl, we simply quote our previous results: the time when 
the fluctuation mode $k$ crosses the Hubble radius is  
\eqn\cor{\tau(k) = l_s^2 k \left({k \over k_c}\right)^{2 \over n-1}
\left(1+{n(2n-5) \over (n-1)(2n-1)}
\left({k_c \over k} \right)^{4 \over n-1} \right),
}
the power spectrum reads approximately 
\eqn\lak{P (k) = B k^{-{2 \over n-1}}
\left(1- {4n^2(n-2)(2n+1) \over (n+1)^2 (n-1) (2n-1)} 
\left({k_c \over k} \right)^{4 \over n-1}  \right),
} 
where
\eqn\laka{B = \left({n (2n-1) \over (n+1)^2} \right)^{n \over n-1}
{n \over 8 \pi^2}
\left({l_p \over l}\right)^2 l^{- {2 \over n-1}}.
}
Requiring $P (k=0.05 Mpc^{-1}) \sim B k^{-2/(n-1)} 
\sim 2 \times 10^{-9}$ for $n=13.17$ 
and $k_c =1.65 \times 10^{-5}Mpc^{-1}$ (these values are from \hl), 
we obtain $l \sim 1.19 \times 10^{-24}$cm and 
$l_s \sim 4.35 \times 10^{-29}$cm . Our result on $l_s$ is one order
of magnitude smaller than the one obtained in \tmb.
The above result is obtained using our old results \hl, we shall
determine all these parameters using numerical method in the
following, the result will be slightly different.

Here we notice that we expand the power spectrum for UV modes in terms 
$\left(k_c/k\right)^{4 \over n-1}$ which equals 0.071 for $k=0.05Mpc^{-1}$ 
and 0.21 for $k=0.002Mpc^{-1}$. We can certainly trust 
the results for $k=0.05Mpc^{-1}$. So in order to get better
results for smaller $k$ and to fit the experimental data in general, 
we need to compute the power spectrum numerically. 

In the numerical solution, we need to find the $k_t$ first. 
If a fluctuation mode $k$ was created inside the Hubble radius, the time 
$\tau(k)$ when it crossed the Hubble radius can be determined by 
\eqn\cross{f(\tau,k) = {z_k'' \over z_k} - k^2 = 0,}
here the primes denote derivative with respect to $\eta$. 
We already defined the critical $k_t$ above, it satisfies the above
relation as well as \create. After determining $k_t$, we compute the power
spectrum  for the IR modes and the UV modes ($k < k_t$ and $k > k_t$) separately. 

For the IR modes, using equation \create, we have
\eqn\dtkc{{d \tau \over d k} 
=l_s^2  
\left(1+{n+1 \over n}\left({k \over k_s}\right)^{2 \over n} \right)
\left(1+\left({k \over k_s} \right)^{2 \over n} \right)^{-{1 \over2}}.
}
For the second case, using equation \cross, 
\eqn\dtk{{d \tau \over d k}
= - {\p f(\tau,k) / \p k \over \p f(\tau,k) / \p \tau}.}
This formula is too complex to be written down here. Armed with these
formulas, the spectral index and its running can be expressed as 
\eqn\index{n_s = {d \ln P(k) \over d \ln k} 
= {k \over P(k)} \left({\p P(k) \over \p k}+{d \tau \over d k} 
{\p P(k) \over \p \tau}\right),}
\eqn\running{{d n_s \over d \ln k }
= {k} \left({\p n_s \over \p k}+{d \tau \over d k} 
{\p n_s \over \p \tau}\right),}
where $P(k)$ has been given in equation \power.

There are three parameters ($l$, string scale $l_s$ and the power n) 
in the noncommutative power-law inflation. The main 
constraints from WMAP \wmap\ are the amplitude of the power spectrum 
$P_k(k=0.002 Mpc^{-1})= 2.09 \times 10^{-9}$, the spectral index 
$n_s(k=0.05 Mpc^{-1})=0.93^{+0.02}_{-0.03}$, 
$n_s(k=0.002 Mpc^{-1})=1.20^{+0.12}_{-0.11}$, 
and the running of the spectral index 
$d n_s / d \ln k (k=0.05 Mpc^{-1})= -0.031^{+0.016}_{-0.017}$, 
$d n_s / d \ln k (k=0.002 Mpc^{-1})= -0.077^{+0.050}_{-0.052}$. 
We will not try to tune three parameters to best fit the experiments data. 
Rather, we fit the data by picking  $n=12$, $l = 2.15 \times 10^{-24}$cm and 
$l_s = 3.86 \times 10^{-29}$cm.  The critical wave number is then 
$k_t = 11.9 \times 10^{-4} Mpc^{-1}$ (slightly larger than 
$H_0=4.6\times 10^{-4} Mpc^{-1}$).  
We show the numerical results in Figure 1, 2 and 3. 
If we dial $n$ downward, the running of the spectral index becomes larger.
The effects of noncommutative spacetime in a general inflation model
will be discussed elsewhere, and it is found that one of the slow-roll
parameter $\epsilon$ and the noncommutative parameter control the running of the
spectral index, a larger $\epsilon$ results in a larger running.

\bigskip
{\vbox{{\epsfxsize=8cm
        \nobreak
    \centerline{\epsfbox{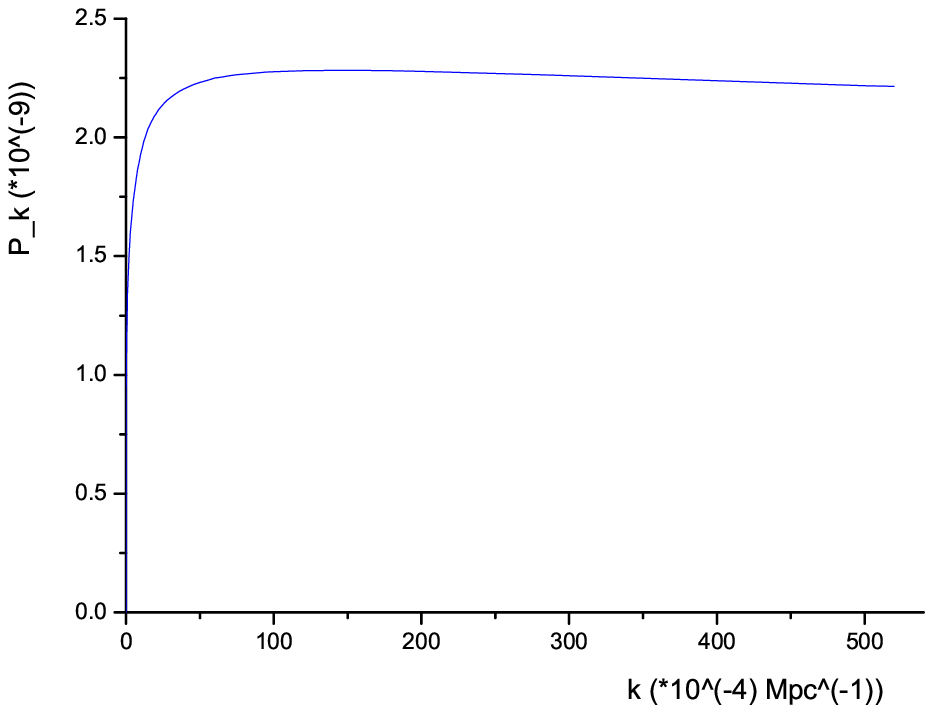}}
        \nobreak\bigskip
    {\raggedright\it \vbox{
{\bf Figure 1.}
{\it $P_k$ is the amplitude of the power spectrum and k is the comoving
mode.  }
 }}}}
    \bigskip}

\bigskip 
{\vbox{{\epsfxsize=8cm 
        \nobreak 
    \centerline{\epsfbox{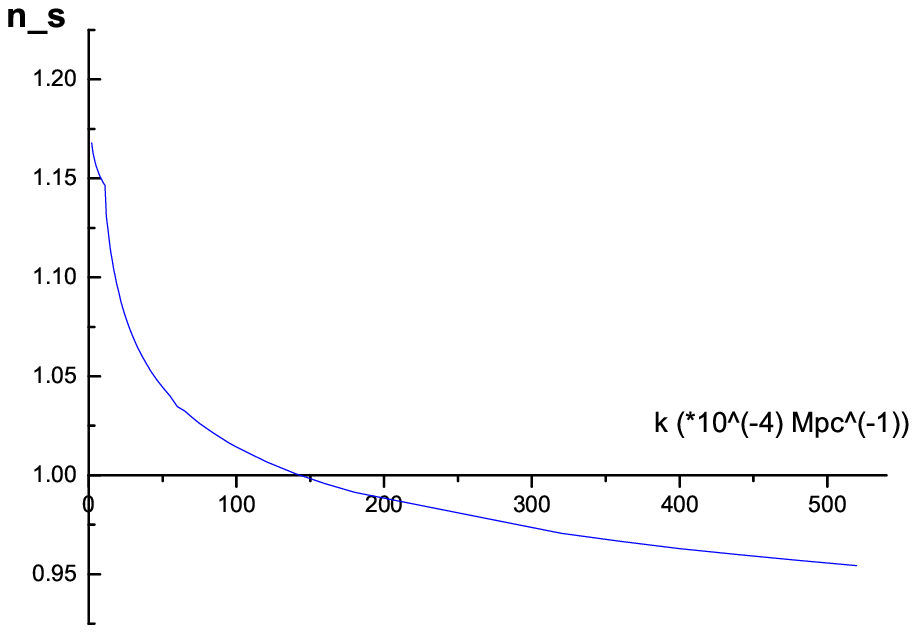}} 
        \nobreak\bigskip 
    {\raggedright\it \vbox{ 
{\bf Figure 2.} 
{\it $n_s$ is the spectral index and k is the comoving mode.  } 
 }}}} 
    \bigskip}  

\bigskip
{\vbox{{\epsfxsize=8cm
        \nobreak
    \centerline{\epsfbox{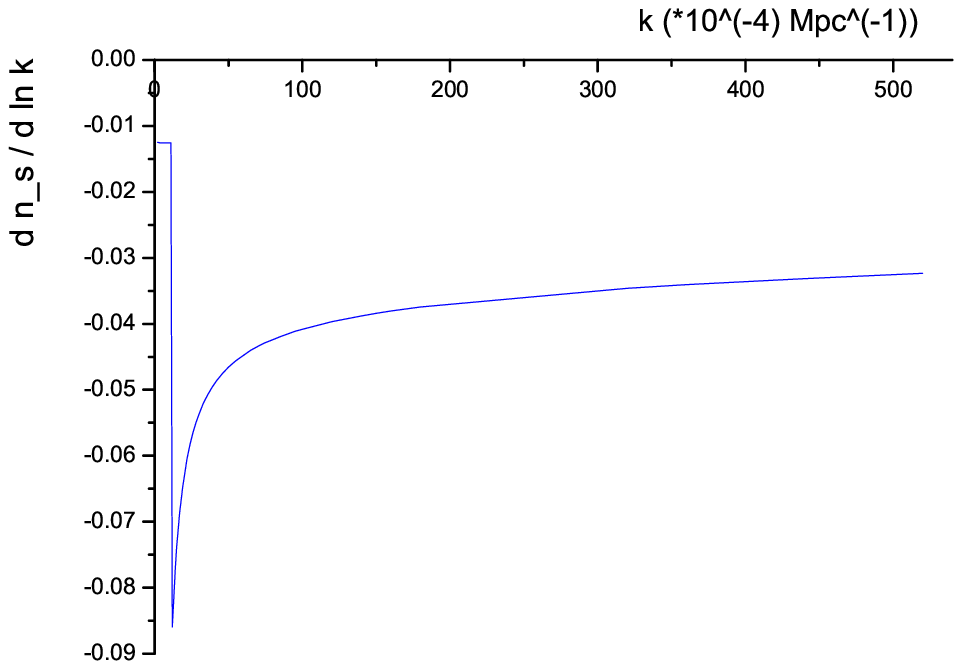}}
        \nobreak\bigskip
    {\raggedright\it \vbox{
{\bf Figure 3.}
{\it $d n_s / d \ln k$ is the running of the spectral index and k is the 
comoving mode. }
 }}}}
    \bigskip}

>From Figure 2 and 3, we read off 
\eqn\res{n_s = 0.956, \quad {d n_s \over d \ln k}=-0.0325 \quad \hbox{at}
\quad k=0.05 Mpc^{-1};}
and 
\eqn\rest{n_s= 1.093, \quad {d n_s \over d \ln k}=-0.0632 \quad
\hbox{at} \quad k=0.002 Mpc^{-1}. }
These results fall within the error bars of the WMAP data. One can certainly
improve these results, but the main conclusion will not change: the noncommutative
power-law inflation model can fit the tentative running spectral index data
of WMAP, and the string scale so determined is about $10^{-29}$cm, 4 order of
magnitude larger than the Planck length.

Although it is not easy to explain the great difference between the Planck scale
and the string scale in the perturbative string theory (the Planck scale is
related to the string scale by $l_p=l_s\sqrt{\alpha_{\hbox{GUT}}}/2$ in
the perturbative heterotic string theory), it is not hard to tune the string 
scale all the way up to $10^{-18}$cm in type I compactifications and nonperturbative
heterotic string theory. Although our result is rather tentative, it offers
some promise to determine the string scale experimentally. Doubtlessly, 
further experimental data will bring forward more surprises, and hopefully
consolidate the existing tentative results on the running spectral index.

Finally, our result on the CMB angular power spectrum is shown in Figure 4. We see
that 
the noncommutative effects indeed suppresses the low multipoles, but not to
the degree required by the WMAP results.

\bigskip
{\vbox{{\epsfxsize=8cm
        \nobreak
    \centerline{\epsfbox{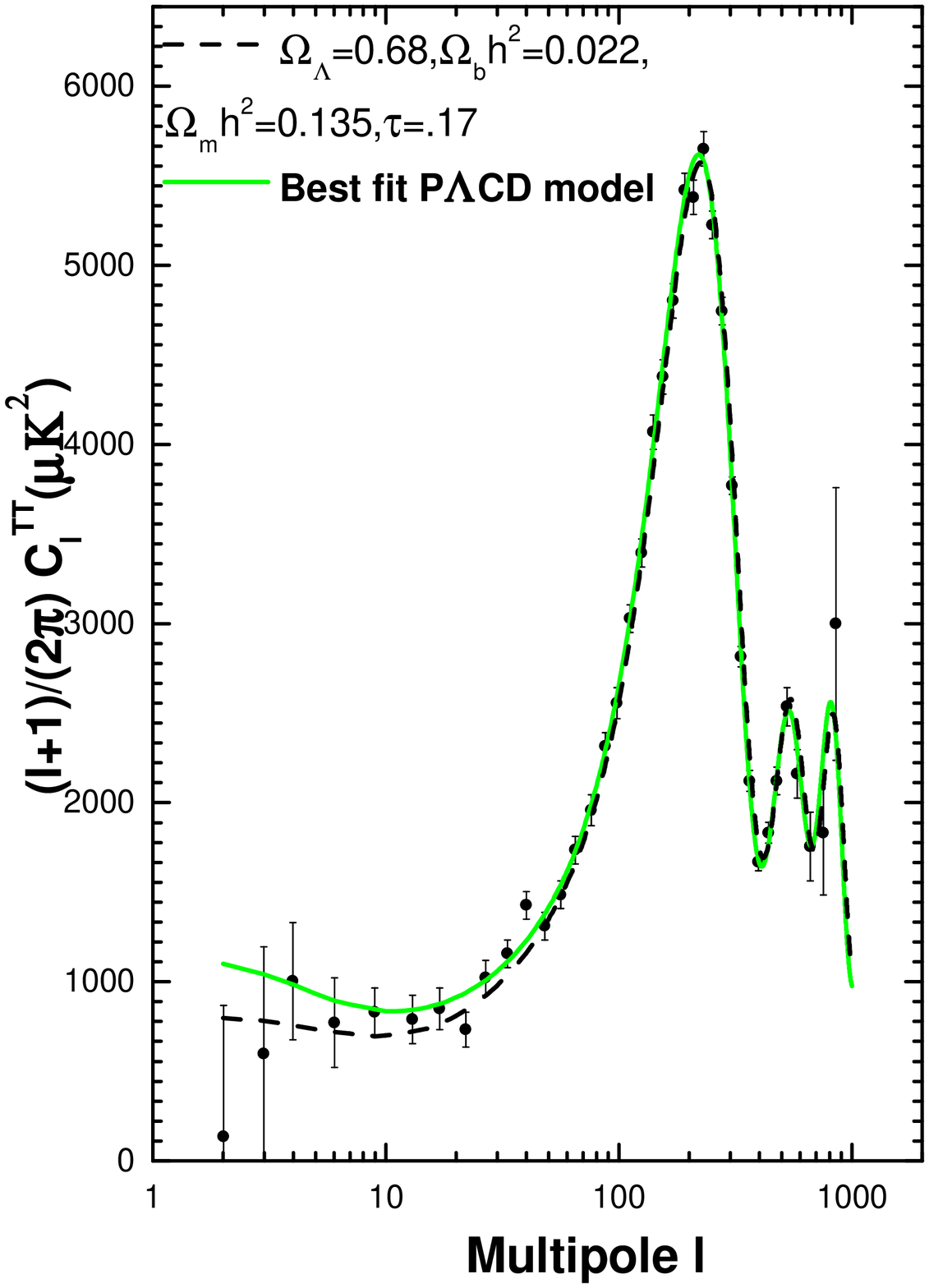}}
        \nobreak\bigskip
    {\raggedright\it \vbox{
{\bf Figure 4.}
{\it The CMB angular power spectrum of the spacetime noncommutative 
power-law inflation. The dashing line is our model and the green solid
line is the Best fit P$\Lambda$CDM model, P stands for power-law. }
 }}}}
    \bigskip}

We emphasize that our analysis and results are different from those in \tmb\
in some technical details.

The results presented here, as in \hl\ and \tmb, are exclusively on the
noncommutative power-law inflation. A general analysis on noncommutative
inflation based on the scheme of \bh\ will be presented elsewhere. Also,
it would be interesting to study the general consequences of noncommutative
spacetime for inflation without resorting to a definite scheme.

We conclude that the running of the spectral index of the CMB power spectrum
may strongly indicate that new physics such as noncommutative spacetime in
string theory operated during inflation, and that if the deficit of the
power for low multipoles is true, still more new physics is required. We are
eagerly waiting for new data from WMAP which may or may not consolidate
both of these surprising features.

\bigskip 

Acknowledgments.
We would are grateful to Y. S. Piao and 
X. M. Zhang for useful discussions, especially to B. Feng for
discussions and help in numerical calculation.
This work was supported by a
``Hundred People Project'' grant of Academia Sinica
and an outstanding young investigator award of NSF of China.

\listrefs 
\end